\documentclass{conm-p-l}

\theoremstyle{definition}

\theoremstyle{remark}

\numberwithin{equation}{section}

\def\be{\begin{eqnarray}}
\def\ee{\end{eqnarray}}
\def\bee{\begin{eqnarray*}}
\def\eee{\end{eqnarray*}}

\def\ds{\displaystyle}
 
\def\bra{\langle}
\def\ket{\rangle}
\def\kb{ \ket \bra }

\def\ot{\otimes}

\def\raw{\rightarrow}

\def\half{{\textstyle \frac{1}{2}}}

\begin{document}

\title{Comments on Adiabatic Quantum Algorithms}

\author{Mary Beth Ruskai}
\address{Department of Mathematics, University of Massachusetts  Lowell, 
Lowell,  MA  01854 USA}
\email{MaryBeth\_Ruskai@uml.edu}
\thanks{Partially supported  by 
 the National Security Agency (NSA) and
 Advanced Research and Development Activity (ARDA) under
Army Research Office (ARO) contract number
   DAAG55-98-1-0374 and by the National Science
        Foundation under Grant number DMS-0074566.}




\date{}

\copyrightinfo{2002}{by author. Reproduction of this article,
  in its entirety,  is permitted for non-commercial
  purposes.}

\begin{abstract}
Recently a method for adiabatic quantum computation has been
proposed and there has been considerable speculation about its
efficiency for NP-complete problems.   Heuristic arguments 
in its favor are based on the unproven assumption of an
eigenvalue gap.
  We show that, even without the assumption of an eigenvalue gap,
other standard arguments can be used to show that a large class
of Hamiltonians proposed for adiabatic quantum computation
 have unique ground states.

 We also discuss some of the issues which arise in trying to
analyze the behavior of the eigenvalue gap.   In particular,
we propose several mechanisms for modifying the final Hamiltonian
to perform an adiabatic search with efficiency comparable to
that for 3-SAT.   We also propose the use of randomly
defined final Hamiltonians as a mechanism for analyzing the
generic spectral behavior of the interpolating Hamiltonians
associated with problems which lack sufficent structure to be
amenable to efficient classical algorithms.

\end{abstract}

\maketitle

\section{Introduction}

Recently there has been considerable interest in a proposed scheme
for adiabatic quantum computation \cite{FGG1,FGG2} and speculation that
it may
even provide a mechanism for efficient solution of hard problems.
Both the validity of the adiabatic theorem and the arguments 
\cite{CFP,FGG1} for its
efficiency depend on the existence of an eigenvalue gap.
However, the existence of such a gap has not been proven.
It has been conjectured \cite{FGG2}  on the basis of numerical
simulations and the so-called non-crossing rule.

The adiabatic quantum algorithm is designed to take the
ground state of an initial Hamiltonian $H_0$ to that of a
final Hamiltonian $H_1$ using a
linear interpolating Hamiltonian of the form
$ H(s) = (1-s) H_0 + s H_1$.  The quantum adiabatic theorem
\cite{HaJy1,KatoAd,LL,Thg} can be used to show  
that the efficiency of the adiabatic approximation is
$O(1/g_{\min}^2)$ where $g_{\min}$
denotes the minimum energy gap $g(s) = E_1(s) - E_0(s)$
between the ground and first
excited states of $H(s)$.  Thus the efficiency of adiabatic
quantum computation depends on how rapidly the eigenvalue
gap decreases as the size of the problem increases.  The
purpose of this note is not to resolve this question, but
to discuss some of the issues that arise.


 Farhi, et al \cite{FGG1,FGG2}
(hereafter referred to as FGG) typically use an initial
Hamiltonian  of the form 
\be  \label{eq:Ham0}
H_0 = \half \sum_j a_j \, [\sigma_x(j) + I] \equiv \half  \sum_j  a_j
     I \ot I \ldots \ot [\sigma_x + I] \ldots  \ot I
\ee
with $a_j$ a non-negative integer. The eigenstates of $H_0$
are products of eigenstates of $\sigma_x$.
The so-called  ``computational basis'' $|k_1 \ldots k_n \ket$
consists of products of eigenstates of the Pauli matrix
$\sigma_z$.  (However, it is customary to identify each $k_i$
with an element of $Z_2 = \{ 0,1 \}$ rather than with the
usual eigenvalues of $\pm 1$ or, equivalently, replacing
$\sigma_z$ by $\half [\sigma_z +I]$ as has been done above.) 
In the computational basis, the eigenstates of $H_0$ have
the form 
$ 2^{-n/2} \sum_{k_1 \ldots k_n} \pm 1 \,  |k_1 \ldots k_n \ket \,$
(with all  signs  $+1$ for the ground state).   FGG
define a non-negative final Hamiltonian $H_1$ which
is diagonal in the computational basis and has the solution of 
some problem as its ground state, e.g., they 
encode the  problem known as ``3-SAT'' in the computational basis.
We note here only that this encoding results in a Hamiltonian
of the form  
\be \label{eq:Hamfin}
H_1 = \sum_{k_1 \ldots k_n} E_{k_1 \ldots k_n}
              |k_1 \ldots k_n \kb k_1 \ldots k_n |
\ee
where the ground state energy is zero and the other 
$E_{k_1 \ldots k_n}$ are positive integers with an upper
 bound that is $O(n^3)$.


First, we point out that the non-crossing rule is completely
inadequate for the purpose of establishing a gap.   
There are realistic models of physical systems
which do exhibit crossings, despite the absence of a symmetry
common to both the initial and final Hamiltonians $H_0$ and $H_1$.
The most well-known such system is $H_2^+$, the hydrogen
molecule ion \cite{Judd}.  Another example is the Hubbard
model for benzene \cite{HL}.   The non-crossing rule and the limitations
on its applicability to adiabatic quantum computation are discussed
in Appendix~\ref{sect:Non}.

However, the most one could hope to gain from the non-crossing rule
is uniqueness of the ground state.   Fortunately, a standard
argument based on the Perron-Frobenius theorem \cite{HJ1,MM}
suffices for that purpose.  This argument is sketched in the
Section~\ref{sect:PF}, but does not address the more fundamental issue
of the size of the resulting eigenvalue gap.

Some insight into the issues raised by adiabatic computation
can be obtained by considering various strategies for
replacing FGG's final Hamiltonian $H_1$ by a modification
whose ground state is the solution of Grover's search problem.
We argue that if their algorithm is sufficiently
robust to solve an NP problem in polynomial time
$O(N^p)$, then a modification should be able to perform a 
successful search of an unordered list of $M$ items in
$O([\log M]^p)$ time, violating the conventional wisdom
that Grover's algorithm (which requires $O(\sqrt{M})$ time) is
optimal.   A model problem  suggested by FGG shows that this 
simplistic expectation need not hold; however,
their example also suggests that gaps are associated with
the presence of a symmetry, not its absence.

The issues raised here involve questions in several subfields
fields of physics and computer science.   In order  to
make this note accessible to people with diverse backgrounds
ranging from Schr\"odinger operator theory to computer science,
two appendices are included --- one on Grover's algorithm and 
one on the non-crossing rule.


\section{Uniqueness of the Ground State} \label{sect:PF}

Although proof of an eigenvalue gap is  likely to be difficult,
proving that systems of the type considered by FGG have
a {\em unique} ground state is easier.   It relies on a standard
argument widely used \cite{RS4} to prove uniqueness (and positivity)
of the ground state in a variety of systems, including quantum 
lattice models.   There is no particular
originality in the argument given below. 
 We present it only in the hope of
 clarifying some issues, particularly the  distinction between the
 uniqueness of the ground state and the existence of a
lower bound on the size of the resulting eigenvalue gap.

 The ground state of the initial Hamiltonian (\ref{eq:Ham0})
  is easily seen to be unique and  consists of products of ground
states of $\sigma_x(k)$.   However, we wish to transform $H_0$
to the computational basis 
of products  $| k_1 k_2 \ldots k_n \ket$ of eigenstates of $\sigma_z(k)$.
 This is easily achieved using
tensor products of the Hadamard transform.   Instead of
examining $H_0$ itself, we consider the operator $F  = e^{-H_0}$.
The ground state of $H_0$ is 
the eigenfunction corresponding to the largest 
eigenvalue of the matrix 
\be
F = e^{-H_0} = \bigotimes_{k=1}^n e^{-a_k  [\sigma_x(k) + I]/2} = 
 B_1 \ot B_2  \ldots \ot B_n
\ee
where 
\be
  B_k & = & \half e^{a_k/2} 
  \left( \begin{array}{cc} 1 & 1 \\ 1 & -1  \end{array} \right) 
\left( \begin{array}{cc}  e^{a_k/2} & 0 \\ 0 & e^{-a_k/2} 
       \end{array} \right)  
       \left( \begin{array}{cc} 1 & 1 \\ 1 & -1 \end{array} \right)   
     \nonumber \\
 & = & e^{a_k/2}
     \left( \begin{array}{cc} \cosh \frac{a_k}{2} & \sinh \frac{a_k}{2} 
       \\  \sinh \frac{a_k}{2} & \cosh \frac{a_k}{2}  \end{array} \right) 
\ee
Now, since all elements of each $B_k$ are strictly
positive, their tensor product $F = e^{-H_0}$ also has strictly
 positive
elements.  Hence, by the Perron-Frobenius theorem \cite{HJ1,MM}
the largest eigenvalue of $F$, and the ground state of $H_0$,
is unique.  

We would like to know that the ground state remains unique for
a Hamiltonian of the form $H = (1-s) H_0 + s H_1$.  For this,
it suffices that $H_1$ is diagonal (in the computational basis)
and has a unique ground state.  By the Lie-Trotter formula
\be
  e^{-H} = \lim_{m \raw \infty} 
    \left(e^{-\frac{s}{m} H_1} e^{-\frac{1-s}{m} H_0}  \right)^m
\ee
The effect of the diagonal matrix $e^{-\frac{s}{m} H_1}$ is
simply to multiply each row of $e^{-\frac{1-s}{m} H_0}$ by a
positive number of the form $e^{-\lambda_j t/m}$.  Hence the
product $e^{-\frac{s}{m} H_1} e^{-\frac{1-s}{m} H_0}$ also has
positive elements and so does its $m$-th power.  
(Moreover, because $e^{-\lambda_j s/m} \raw 1$ as $m \raw \infty$,
these positive elements do not become zero in the limit.) 
Thus, one
can again apply the Perron-Frobenius theorem to conclude that
the largest eigenvalue of $e^{-H}$ is unique and, hence, $H$
has a unique ground state if $0 \leq s < 1$.  (The argument
 breaks down at
$s = 1$ because $e^{-\frac{1-s}{n} H_0} = I$ no longer has
strictly positive elements off the diagonal and is completely
decomposable.)  

Note that, even if $H_1$ has a degenerate ground state, 
the interpolating Hamiltonian
$H = (1-s) H_0 + t H_1$ will still have a unique
ground state for all $0 \leq s < 1$; however, the difference
between the
two lowest eigenvalues must $\raw 0$ as $s \raw 1$.  Thus,
uniqueness of the ground state is a very different matter
from an eigenvalue gap of minimal size.  Indeed, the uniqueness
argument above holds for models \cite{FGG1}, such as an adiabatic
search for which the gap 
can be shown to decrease exponentially.

In addition to uniqueness, the Perron-Frobenius theorem implies
that the ground state has the form $\sum_k c_k | k_1 k_2 \ldots k_n \ket$
with strictly positive coefficients $c_k > 0$ in the computational
basis for all $s \in [0,1)$.  In the limit as $s \raw 1$ all but
one of these $c_k \raw 0$.


\section{Adiabatic Search Algorithms} \label{sect:ad.srch}

As described in Appendix~\ref{Sect:Grov},
Grover's \cite{Grov1,Grov2,NC} algorithm is
designed to efficiently locate an unknown but identifiable target
via the use  of  a unitary operator $G$ which can be
written as  $G= e^{i \pi A}$ where
\be
 A \, | k_1 \ldots k_n \ket = \left\{ \begin{array}{cl}
 0    & 
   \hbox{if}~  (k_1 \ldots k_n) =  (t_1 \ldots t_n)  \\
     | k_1 \ldots k_n \ket   & \hbox{otherwise} 
\end{array}   \right.
\ee
One can easily implement an adiabatic search by 
choosing for the final Hamiltonian $H_1$ in the FGG algorithm 
the Grover generator, $A$, above.
The adiabatic evolution will take the ground state of $H_0$, namely,
$| \psi_0 \ket \equiv
2^{-n/2} \sum_{k_1 \ldots k_n}   | k_1 \ldots k_n \ket$
 to the ground state of $H_1 = A$, namely $| t_1 \ldots t_n \ket$.
However, FGG have shown that this process takes exponential
time.    The analysis can be simplified \cite{WMV} by modifying the
initial Hamiltonian $H_0$ to reduce the analysis to a two-dimensional
problem.

The reduction to a two-dimensional problem, which plays a
critical role in Grover's algorithm, is associated with
 a $(2^n \!- \!1)$-fold
degeneracy in the adiabatic Hamiltonian $H_0$ and $H_1$.
However, this is not at all essential for the success of
an adiabatic search.  All that is needed is that the ground
state of $H_1$ be the target state $| t_1 \ldots t_n \ket$.
This suggests that one try to modify $H_1$ so that its
ground state is $| t_1 \ldots t_n \ket$, but the eigenvalue
distribution of its excited states is similar to that of
a final Hamiltonian known to have a gap.

Suppose that a problem is known to have an efficient solution
encoded in the final Hamiltonian $H_1$.
Then setting $H_2 = G H_1 $,  yields a Hamiltonian 
identical to $H_1$ except that the eigenvalue 
associated with the target
state $| t_1 \ldots t_n \ket$  is multiplied by $-1$.  
(Because $G$ and $H_1$ are both diagonal in the
computational basis, they commute and $H_2$ is self-adjoint.)
Because $H_1$ was defined to be
 non-negative, $H_2$ has exactly one negative 
eigenvalue so that its ground state is now the target state
$| t_1 \ldots t_n \ket$.    
Applying the adiabatic algorithm to the modified interpolating
Hamiltonian $ H_2(s) = (1-s) H_0 + s H_2 $
should take the ground state of $H_0$ to the target state.
Moreover, the {\em only} effect on the final Hamiltonian is 
to move one excited state below the previous ground state, 
without decreasing the final energy gap $g(1)$.
A similar  Hamiltonian which is
non-negative could be constructed the replacement
\bee
  E_{k_1 \ldots k_n}  \raw & 0  &{\rm if} ~ k_1 \ldots k_n = t_1 \ldots t_n \\
  E_{k_1 \ldots k_n}  \raw  & E_{k_1 \ldots k_n} + 1 & {\rm otherwise} .
\eee
in (\ref{eq:Hamfin}).   In either case, one would
 not generally expect these modifications to
significantly affect $g_{\min}(s)$,
in which case the adiabatic search would
be as efficient as the solution of the problem
encoded in $H_1$.  

Thus, if Farhi, et al's projection of an efficiency of
$O(n^p)$ is correct for the problem encoded in $H_1$, 
then one would expect an efficiency of
$O([\log N]^p)$ for the adiabatic search of a list of
$N = 2^n$ items described above.    
However, this would imply that a
quantum computer could search an unordered list in time
$O([\log N]^2)$, contradicting the conventional wisdom that a speed-up
greater than $O(\sqrt{N})$ is not possible \cite{BBBV,BBHT,NC,Z}.
This does not necessarily imply a contradiction.  The proofs 
that $O(\sqrt{N})$ is optimal depend on assumptions about
the nature of the ``oracle query'' used in the search.   
However, van Dam, Mosca and Vazirani \cite{WMV} have observed
that the
encoding of solutions of other problems, such as 3-SAT,
 in $H_1$ implicitly
assume the ability of the computer to perform more general
queries.  Thus, standard complexity query arguments can not
rule out the possibility of polynomial time algorithms.

After seeing a preliminary version
of this manuscript, FGG \cite{FG} pointed out that  the 
change from $H_1$ to $H_2$ described above
can have a dramatic effect on the gap.  We describe their
example in the next section.   

It should also be noted that our  expectation that the change
from $H_1$ to $H_2$ will not decrease the minimum gap is {\em not}
based on the presumption that one is a small perturbation of the
other.   On the contrary, (as FGG  \cite{FG} emphasized) the two
Hamiltonians differ by a multiple of a projection, which can have
a significant effect on the spectrum.  Indeed, it is essential
to our strategy to effect such a change in the {\em final}
Hamiltonian.   However, unless this also induces a change in the
structure of the problem, such as a symmetry-breaking, this need
not affect the generic behavior of the spectra of the 
interpolating Hamiltonian; in particular, it need not lead to
an avoided crossing of the two lowest levels.


\section{A Separable Model}  \label{sect:sepmod}

Let $H_1 = \half \sum_j [ \sigma_z(j) + I]$ and set $a_j = 1$
in $H_0$. Then the interpolating  Hamiltonian  becomes
\be
H(s) =  H_0 + s(H_1 - H_0) =
 \half  \sum_j \Big[ (1-s) \sigma_x(j) + s  \sigma_z(j) + I \Big]
\ee
This system is separable, and exactly solvable.  Since 
$[ (1-s) \sigma_x(j) + s  \sigma_z(j) + I \Big]$ 
\linebreak has eigenvalues 
$\half ( 1 \pm \sqrt{1 - 2s +2s^2}\,)$, the ground state of $H(s)$ 
has energy  \linebreak $\frac{n}{2}( 1 - \sqrt{1 - 2s +2s^2}) $ and the
gap $g(s) = \sqrt{1 - 2s +2s^2}$
 is independent of $n$ with $g_{\min} = \frac{1}{\sqrt{2}}$.  
The system also has
a high level of symmetry, since $H(s)$ commutes with elements
of the symmetric group $S_n$.  In fact, there is an additional
accidental degeneracy so that the $(k+1)$-st eigenvalue is $k$ with
a degeneracy of $\left( \begin{array}{c} n \\ k \end{array} \right)$ 
 for $k = 0,1, 2 \ldots n$.

Now if  $H_1 $ is replaced by $H_2 = GH_1$, the symmetry is
broken.  However, there is still some symmetry and a high
level of degeneracy.  Essentially, only one state from each of
the $(n+1)$ non-degenerate levels is affected and the problem reduces
to an  $(n+1)$-dimensional one which can be analyzed explicitly
and shown to have an exponentially decreasing gap.  In effect,
an eigenvalue can be associated with the target state and must
 ``cross'' the levels lying below it to reach the bottom
of the spectrum when $s = 1$.  (A more detailed
examination shows that the levels become exponentially close
and then bounce away with the target information transmitted
to the lower level).

Although this shows that the argument sketched in 
Section~\ref{sect:ad.srch} above cannot be made rigorous, this
model is not generic.   The
exponentially decreasing gap is the result of a symmetry
breaking which should not occur when the Hamiltonian $H_1$
has no symmetry to begin with.

\section{Discussion}


Farhi, et al \cite {FGG1} analyze several other models for which
the gap behavior can be calculated explicitly and shown to
decrease slowly (i.e, polynomial in $n$).  However, as they
point out, these models
all have a high level or symmetry or structure which would lead
to efficient classical algorithms.  In some cases
symmetry allows a high level of degeneracy which permits
one to squeeze $2^n$ states into $[0,O(n^p)]$ without forcing an
exponentially decreasing gap.  

In adiabatic computation the typical choices for initial
and final Hamiltonians have spectra with high 
degeneracies and consist of positive integers in
 a range that is polynomial in $n$.  In general, the
interpolating Hamiltonian $H(s)$ breaks these degeneracies
and must  squeeze $2^n$ distinct eigenvalues into a range of
the form $[0, O(n^p)]$.  Thus, most of them must be 
exponentially close.

It seem that a polynomial gap is more likely to be associated
with the presence of symmetry, which allows high degeneracies,
 than with its absence.

For excited states, it is irrelevant whether or not the 
observed mergings   are ``avoided crossings'' or true
crossings.   The simultaneous coalescence of a large number of
excited states is essential to  the algorithm.
The states of the initial and final Hamiltonians are both
product states, those of the initial Hamiltonian are products
of eigenstates of $\sigma_x$, while those of the final
Hamiltonian  are products of eigenstates of $\sigma_z$
(and thus elements of the so-called ``computational'' basis).
Therefore, the final ground state is always an
evenly weighted superposition of {\em all} eigenstates of the
initial Hamiltonian.   Unlike standard applications of the
adiabatic theorem, in which the main contribution comes from
a few low-lying states, {\em all} of the excited states must
contribute to the first-order correction.   To do this, these 
higher excited states must get close in some sense.
  Fortunately, a quantum computer can make
this first order correction efficiently, mixing in all $2^n$
excited states, and this is where the method gets its potential power.
However, in order that low order perturbation theory suffice,
 it is essential
that the gap between the ground and first excited state not 
decrease too rapidly as $n$ increases. 

The non-crossing rule, which is discussed in Appendix~\ref{sect:Non}
is based on the belief that ``accidents'' are extremely rare
so that such phenomena as persistent degeneracy, or level
crossings do {\em not} occur without some underlying physical
phenomenon (such as a symmetry) with implications for the
associated mathematical model.  This viewpoint would suggest that
if the lowest gap is to decrease only polynomially when the other
eigenvalues are getting exponentially close, there must be
some physical mechanism keeping them apart.   We are  
skeptical that such a mechanism can be found for 
Hamiltonians which encode the solution of problems which
do not have enough structure to yield efficient classical
solutions.

This raises another question.  Is the spectrum
of the interpolating Hamiltonian sensitive to the association
of particular eigenvalues with  particular eigenstates in
the final Hamiltonian, or is it primarily dependent on the
eigenvalue distribution?  For Hamiltonians with a good deal of
structure, the first situation clearly holds, and FGG have
observed \cite{FGG1,FGG3} 
that the behavior of the eigenvalue gap can depend
critically on the choice of initial Hamiltonian.   However,
one expects Hamiltonians which encode  solutions to typical
instances of  classically intractable problems, to lack 
the structure needed for this sensitivity.

Because of the difficulty in analyzing the behavior of the
gap in problems without structure, it may be worth considering
a randomly defined Hamiltonian, i.e., let the energy in 
(\ref{eq:Hamfin}) have the form 
$ E_{k_1 \ldots k_n} = f(k_1 \ldots k_n)$ where $f$ is a
suitable random process.  One can then ask if there is
a sense in which the  eigenvalue gap is ``almost always"
exponentially decreasing.   If so, this would suggest that
an exponentially small gap is generic, and likely to occur
for NP-complete problems.   There is already an extensive
literature \cite{AzM,AzS,CL,CFKS} on the spectral behavior of
random Schr\"odinger operators, in which one obtains results
about   typical Hamiltonians with certain properties
rather than one with a fixed potential.   Although the
model Hamiltonians used here have quite a different
structure, and may require the development of new techniques,
this approach seems worth considering.

Indeed, one could even define the final Hamiltonian  for
an adiabatic search by letting
$f$ be  a random variable taking integer 
values in $[1,n]$ and choosing
\bee
  E_{k_1 \ldots k_n}  = & 0  ~~~~~~ &{\rm if} ~ k_1 \ldots k_n = t_1 \ldots t_n \\
  E_{k_1 \ldots k_n}  = & f(k_1 \ldots k_n) & {\rm otherwise} 
\eee
in (\ref{eq:Hamfin}).

\bigskip

\noindent{\bf Acknowledgment:} This note is based on a talk
at the Q-Math8 conference in Mexico in December, 2001.
It is a pleasure to thank the organizers for the opportunity
to present these ideas to an audience of mathematical
physicists in a delightful setting;  to acknowledge
helpful discussions with Professors Charles Bennett, George Hagedorn,
Chris King, John Preskill, Barry Simon,
and Umesh Vazirani; and to thank Professors Edward Farhi and
Sam Gutman for suggesting the model in Section~\ref{sect:sepmod}.
Part of this work was done when the author was visiting the
Institute for Theoretical Physics at the University of
California, Santa Barbara and thereby also partly supported by
the National Science Foundation under Grant PHY-9907949.

 \bigskip

\appendix

\section{Grover's Algorithm} \label{Sect:Grov}

Grover's \cite{Grov1, NC} algorithm is
designed to efficiently locate an unknown but identifiable target
state $|t_1 \ldots t_n \ket$.  This state may be the key denoting
the location of an item in an unsorted list (e.g., the analogue 
of the name in an alphabetized phone book associated with telephone
number one has been given) or a state with a certain verifiable 
property, e.g., a representation of the factors of a given number 
or the solution of some NP-complete problem.

Grover showed how to construct 
 a unitary operator $G$ whose effect is simply to multiply
the unknown target state $| t_1 \ldots t_n \ket$ by $-1$ 
and all others by $+1$.  This operator can then be used to 
construct an algorithm which will find the target state
with probability greater than $\half$ in $O(\sqrt{N})$, i.e., 
$O(2^{n/2})$ time for $N = 2^n$ states.
The operator $G$ is a unitary operator whose action on 
the computational basis is simply
\be
 G \, | k_1 \ldots k_n \ket = \left\{ \begin{array}{rl}
 -   |  k_1 \ldots k_n \ket  & 
   \hbox{if}~  (k_1 \ldots k_n) =  (t_1  \ldots t_n )  \\
     | k_1 \ldots k_n \ket   & \hbox{otherwise} 
\end{array}   \right.
\ee
It is generally assumed that one has an ``oracle'' which can
perform unitary operations to determine whether or not a
state has the desired property and outputs a function $f$ whose
value is $1$ if the answer is yes and $0$ otherwise.  This
oracle is described by the unitary operator which takes
\be
  |k_0 \ket \ot | k_1 \ldots k_n \ket  \mapsto
   |k_0 \oplus f(k_1 \ldots k_n) \ket \ot | k_1 \ldots k_n \ket
\ee
The action of this oracle when the first bit is in the
$\sigma_x$ eigenstate 
$2^{-1/2} \big( | 0 \ket - | 1 \ket$ (for which we use the
somewhat unconventional notation $| 1 \ket_x$) is then
\be
 | 1 \ket_x \ot | k_1 \ldots k_n \ket \mapsto
    e^{i \pi f(k)}  | 1 \ket_x \ot | k_1 \ldots k_n \ket 
\ee
which is exactly the effect of $G$ when the ancillary
initial bit $ | 1 \ket_x $ is omitted.  The power of quantum
computing is then exploited by applying $G$ to 
superpositions of the form 
$\sum_k c_{k_1 \ldots k_n} | k_1 \ldots k_n \ket$ rather
than to the individual states in the computational basis.
The analysis is facilitated by the realization that the
problem can be reduced to a two-dimensional one in
\bee
{\rm span} 
 \Big\{ |t_1 \ldots t_n \ket, \sum_{k_1 \ldots k_n} 
   | k_1 \ldots k_n \ket \Big\}
\eee 
for which $ |t_1 \ldots t_n \ket$ and $\ds{\frac{1}{\sqrt{N-1}} 
   \sum_{k_1 \ldots k_n \neq t_1 \ldots t_n} | k_1 \ldots k_n \ket }$
form an orthonormal basis.

\section{The Non-Crossing Non-Rule} \label{sect:Non}

The so-called ``non-crossing rule'' is one of a number of 
physical principles which arise when symmetry ensures that
a critical term in some expression, such as the leading term in
a perturbation expansion, is zero.  The remaining conditions
needed to obtain a crossing,
transition, etc. are then more easily satisfied.   In the absence of the
canonical conditions, such crossings and transitions are not truly
``forbidden'', but are either rare events which result from an
accidental confluence or the result of {\em  another physical
circumstance} which facilitates the satisfying of certain conditions.

In the case of interest here, the canonical condition for crossing
 is that the Hamiltonians $H_1$ and $H_0$ both commute with the
operators which generate a symmetry group ${\mathcal G}$ (for which 
it is {\em not} necessary that $H_1$ and $H_0$ commute with each other).
The irreducible representations of ${\mathcal G}$ can then be used to
classify the eigenspaces of $H_1$, $H_0$, and the interpolating
Hamiltonian $H(s) = (1-s) H_1 + s H_1$.   The non-crossing
rule asserts that one expects eigenvalues of H(s) to cross only 
if they belong to different irreducible representations.

A similar situation arises in the Born-Oppenheimer approximation
for the hydrogen molecule ion $H_2^+$ in which case the role of
$t$ is replaced by the internuclear distance $R$.   
Standard tables of atomic data, show many instances of
crossings of states in the same  symmetry class, in apparent
violation of the non-crossing rule.  In this case, the paradox
was rather easily resolved by the discovery of a ``hidden symmetry''
which can regarded as a generalization of that responsible for
the well-known ``accidental'' degeneracies of states of different
angular momentum (but same principle quantum number $n$) for hydrogen.
However, the generator of this symmetry group is an operator
(denoted $F(R)$ by Judd \cite{Judd}) which depends on the
internuclear distance $R$.    Although this gives a satisfactory
physical explanation for the phenomena observed, it points out
a difficulty with any attempt to make a mathematically precise
theorem out of the ``non-crossing rule''.  If a crossing exists,
the degeneracy in $H(s)$ always allows the formal construction
of a suitable symmetry group \cite{Heil}.

A less well-known example of violation of the non-crossing rule
occurs in Heilman and Lieb's study \cite{HL} of the Hubbard model 
for benzene.  Moreover, in this case, a rather detailed analysis
\cite{Heil} failed to locate a hidden symmetry, even one dependent on a
parameter.  In addition, their work found persistent degeneracies
unexplained by symmetry.  This is more serious, as it is far less
likely to be an artifact of the numerical methods used.  It is
worth quoting part of their discussion.
\begin{quote}
 The [non-crossing rule] depends[s] crucially on the interpretation
of the word {\em symmetry}.  The conventional meaning is that
of a symmetry group independent of [a parameter] $U$; in this
case the ``proofs'' are false.  \ldots  [If] one allows symmetry
groups that are $U$-dependent, the ``theorems'' are mere 
tautologies, because  \ldots one can always invent, {\em post hoc}
a $U$-dependent group to account for any violations.

One may ask what is wrong with the ``proofs'' quoted above.
The fault lies not in the mathematics {\em per se} but in the
assumptions used to connect the mathematics with the real world:
First, in the natural sciences, two real numbers are never equal
unless there is a physical reason for it; second, that reason
must be the existence of a U-independent symmetry group.

The first assumption has validity, but the second is merely a
confession of ignorance ...
\end{quote}

Although the relevance of the non-crossing rule to adiabatic
quantum computation is questionable, the principles underlying
it are not.  If polynomially decreasing gaps are generic for
systems in which exponentially many eigenvalues are squeezed 
into an interval that increases only polynomially, there must
be a physical reason for this behavior. 

\bigskip


\end{document}